\documentclass[runningheads]{llncs}
\usepackage{booktabs} \usepackage{pgfplots}
\pgfplotsset{compat=newest}
\usepackage{xcolor}
\usepackage{siunitx}
\usepackage{multirow}
\usepackage{xspace}
\usepackage{balance}
\usepackage[font=rm]{caption}
\usepackage{subcaption}
\usepackage{algorithm}
\usepackage[noend]{algpseudocode}
\usepackage{verbatim}
\usepackage{graphicx}
\usepackage{enumitem}
\usepackage{shortvrb}
\usepackage{tikz}
\usepackage{blindtext}
\usepackage[numbers,sectionbib,sort]{natbib}
\makeatletter
\renewcommand{\@biblabel}[1]{#1.}
\makeatother
\renewcommand{\cite}{\citep}
\usepackage[all]{nowidow}

\makeatother
\newcommand{\marco}{{\small{MSMARCO}}}

\newcommand{\Sys}[1]{S_{\mbox{\scriptsize{{#1}}}}}
\newcommand{\Met}[1]{M_{\mbox{\scriptsize{{#1}}}}}
\newcommand{\Jud}[1]{J_{\mbox{\scriptsize{{#1}}}}}
\newcommand{\Top}[1]{T_{\mbox{\scriptsize{{#1}}}}}
\newcommand{\Gol}[1]{\Jud{G}}
\newcommand{\gol}[1]{\hat{g}_{\mbox{\scriptsize{{#1}}}}}

\newcommand{\myparagraph}[1]{\vspace*{-0.25ex}\paragraph*{\normalsize\bf#1.}}
\newcommand{\mycaption}[1]{\caption{{\rm{#1}}}}

\usepackage{xcolor}

\newcommand{\joel}[1]{{\color{orange}{\bf{Joel says:}} \emph{}}}
\newcommand{\alistair}[1]{{\color{purple}{\bf{Alistair says:}} \emph{}}}
\newcommand{\matthias}[1]{{\color{orange}{\bf{Matthias says:}} \emph{}}}

\newcommand{\aftercapspace}{\vspace*{1ex}}

\begin{document}
\title{A Sensitivity Analysis of the MSMARCO Passage
Collection\protect\footnote{The work we report here was carried out in the
period May-August 2021, and was conceived and executed independently
of and concurrently with the complementary work of
{\citet{avyc21arxiv}}.}
}
\author{
Joel Mackenzie\inst{1} \and
Matthias Petri\inst{2}
\and
Alistair Moffat\inst{1}
}
\authorrunning{Mackenzie, Petri, and Moffat \hspace{5cm} arXiv 2021}
\institute{
The University of Melbourne, Australia
\and
Amazon Alexa, United States\\
\email{\{joel.mackenzie,ammoffat\}@unimelb.edu.au, mkp@amazon.com}\\
}
\maketitle              \begin{abstract}
  The recent {\marco} passage retrieval collection has allowed
researchers to develop highly tuned retrieval systems.
One aspect of this data set that makes it distinctive compared to
traditional corpora is that most of the topics only have a single
answer passage marked relevant.
Here we carry out a ``what if'' sensitivity study, asking whether a
set of systems would still have the same relative performance if more
passages per topic were deemed to be ``relevant'', exploring several
mechanisms for identifying sets of passages to be so categorized.
Our results show that, in general, while run scores can vary markedly
if additional plausible passages are presumed to be relevant, the
derived system ordering is relatively {\emph{in}}sensitive to
additional relevance, providing support for the methodology that was
used at the time the {\marco} passage collection was created.
\end{abstract}

\section{Introduction}
\label{sec-intro}

Offline retrieval evaluations make use of {\emph{test collections}},
each of which includes a set of {\emph{documents}}, a set of
{\emph{topics}} (or queries), and a set of relevance judgments (or
{\emph{qrels}}).
A {\emph{run}} is constructed for each combination of system and
topic, and then those runs are scored using an {\emph{effectiveness
metric}}, making use of the qrels for the corresponding topic.
Finally, the run scores are compared across the systems, usually via
a paired (over topics) statistical test {\citep{s10-fntir}}.

The recent {\marco} passage test collection
{\citep{nr+16-nips,cm+21-sigir-marco,cm+21-sigir-trecdl}} differs
from previous test collections, with the ``documents'' short
{\emph{passages}} extracted from larger entities, and with very
sparse qrels.
In particular, there is only a single passage marked as
relevant for the majority of topics, and no passages are marked
non-relevant.
As a result, effectiveness metric values for most runs are drawn from
a small set of distinct values; and systems might risk being deemed
inaccurate if they present equally-attractive, but unjudged, answers
in different orders.

\begin{figure}
\centering
\fbox{
\begin{minipage}{0.9\textwidth}\raggedright\scriptsize
  {\bf{Query:}} how long is super bowl game
  \smallskip

  {\bf{Passage 1:}} A traditional football game is approximately 3
  hours long.
  However, the Super Bowl is approximately 4 hours long from start to
  finish.
  [{\emph{27 more words}}] 
\smallskip

  {\bf{Passage 2:}} However, the Super Bowl is approximately 4
  hours long from start to finish.
  The game is longer due to the lengthened half time show and the
  focus on advertising and commercial breaks.
  [{\emph{82 more words}}]
\smallskip

  {\bf{Passage 3:}} How long does the Super Bowl usually
  last?
  The Super Bowl is typically four hours long.
  The game itself takes about three and a half hours, with a 30
  minute halftime show built in.
[{\emph{63 more words}}]
\end{minipage}}
 \caption{One topic and (extracts of) three passages of {\marco}.
Only the third is marked as being relevant; the other two
are neither relevant nor non-relevant.
\label{fig-threedocs}}
\end{figure}

To illustrate this risk, Figure~\ref{fig-threedocs} shows one of the
{\marco} topics, and the first three passages returned by a standard
BM25 run.
The passage ranked third is the (only) one that has been judged
relevant for this topic, despite the apparent suitability of the
first two.
A system that, perhaps just by chance, had the third answer at rank
one or rank two would have a notably different effectiveness score.
There are many other instances of this effect.

Our goal here is to explore the extent to which unjudged, but
arguably relevant, answers might affect system effectiveness scores,
and also system versus system comparisons.
To do that we develop a range of passage orderings based on
``clairvoyant'' knowledge of the qrel set, including ones that are a
result of fusing multiple held-out systems' runs, and ask a critical
question: if more passages taken from those lists of plausible
candidates are deemed to be relevant, what happens to system scores
and comparative orderings?
Our results show that run scores vary markedly, but that the derived
system ordering is relatively {\emph{in}}sensitive to additional
relevance, providing support for the methodology that was used at the
time the {\marco} passage collection was created.

\section{Experimental Design}
\label{sec-design}

Our goal is to explore score consistency and system ordering stability
as additional passages are assumed to be relevant, augmenting the
set of passages labeled ``relevant'' in the original {\marco} qrels.
The next few paragraphs describe the process used for identifying
plausible candidate passages.

\myparagraph{Notation}

Let $\Sys{}$ be a {\emph{retrieval system}}.
When provided with a query $q$, $\Sys{}$ returns a ranked list of
documents (here, passages) $\Sys{}(q)$.
Further, let $\Met{}$ be an {\emph{effectiveness metric}} which
returns a score derived from a ranking $\Sys{}(q)$ and a set of relevance
judgments for that query, $\Jud{}(q)$.
That is, $\Met{}(\Sys{}(q), \Jud{}(q))$ is the score assigned by
metric $\Met{}$ to system $\Sys{}$ for query $q$, relative to the
judgments $\Jud{}(q)$.
It also convenient to take $q$ as being given, and use the shorthand
$\Met{}(\Sys{}, \Jud{})\equiv \Met{}(\Sys{}(q),\Jud{}(q))$.
Finally, let $\Top{$d$}(L)$ be the first $d$ items in list $L$.
For example, $\Top{$d$}(\Sys{})$ is the first $d$ elements of the
ranking generated by $\Sys{}$ for some query $q$.

\myparagraph{Gold Answers}

The {\marco} qrels establish at least one {\emph{gold answer}} for
each query $q$; we denote $q$'s set of gold answers by $\Gol{}(q)$.
In the {\marco} collection, $|\Gol{}(q)|=1$ for most~$q$; and we
suppose that $\gol{}(q)$ is that passage.
When $|\Gol{}(q)|>1$, we select $\gol{}(q)\in\Gol{}(q)$ as a random
choice.
Given a system, $\Sys{}(\gol{}(q))$ can be computed via a
{\emph{query-by-document}} {\cite{yb+09-wsdm}} mechanism, with
$\gol{}(q)$ likely (but by no means guaranteed) to be the top-ranked
passage.
That is, in the majority of cases,
$\Top{1}(\Sys{}(\gol{}(q)))=\gol{}(q)$; whereas there is no
expectation that $\Top{1}(\Sys{}(q))=\gol{}(q)$.

\myparagraph{Clairvoyant Rankings and Seed Systems}

Our experiments are based on the hypothesis that if some answer $d$
is ``close'' to $\gol{}\in\Gol{}(q)$, then $d$ is also a plausible
candidate for relevance to~$q$ {\cite{ca07-cikm}}.
To quantify closeness, we use the ``query-by-passage'' ordering
$\Sys{}(\gol{})$, and determine the rank at which $d$ occurs.
We can think of $\Sys{}(\gol{})$ as being a {\emph{clairvoyant}}
ranking, since it is derived from knowledge of a relevant passage;
that is, via a relevance feedback loop {\cite{sr03-sigir,
sj04-sigir}}.
We use two different {\emph{seed systems}} to generate those
rankings:
\begin{itemize}

  \item $\Sys{BM}$ is a bag-of-words BM25 run generated using the
  PISA search system {\cite{pisa19-osirrc}} over an Anserini index
  {\cite{yfl18-jdiq}} transferred via the Common Index File Format
  {\cite{lm+20-sigir}} following {\citet{mdgc20-sigir}}.

  \item $\Sys{TCT}$ is the neural TCT-ColBERT-V2-HN+ system described
  by {\citet{lyl21-rl4nlp}}.
  We use Pyserini {\cite{lm+21-sigir}} to conduct brute force
  retrieval via FAISS {\cite{jdj21-ieee}}.

\end{itemize}

\myparagraph{Extrapolated Qrels}

\begin{figure}[t]
  \centering
  \includegraphics[width=0.75\textwidth]{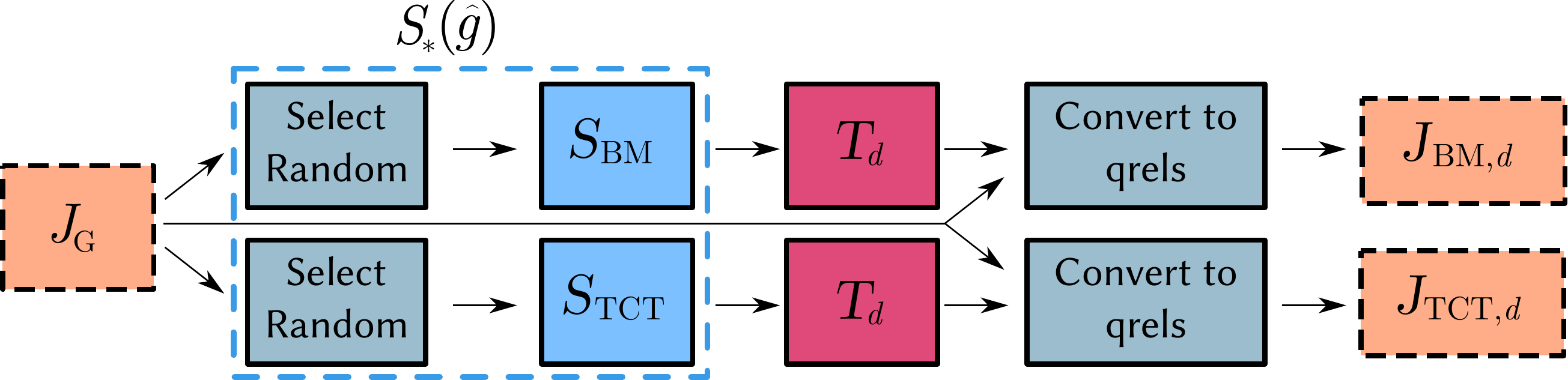}
  \mycaption{Extrapolated judgments.
  A single gold answer from $\Gol{}(q)$ is used as a query to
  generate BM25 and TCT runs.
  The $d$ highest-ranked non-gold passages
  are taken to be relevant, and added to $\Gol{}(q)$.
  \label{fig-basic-pipeline}}
\end{figure}

\begin{figure}[t]
  \centering
  \includegraphics[width=0.75\textwidth]{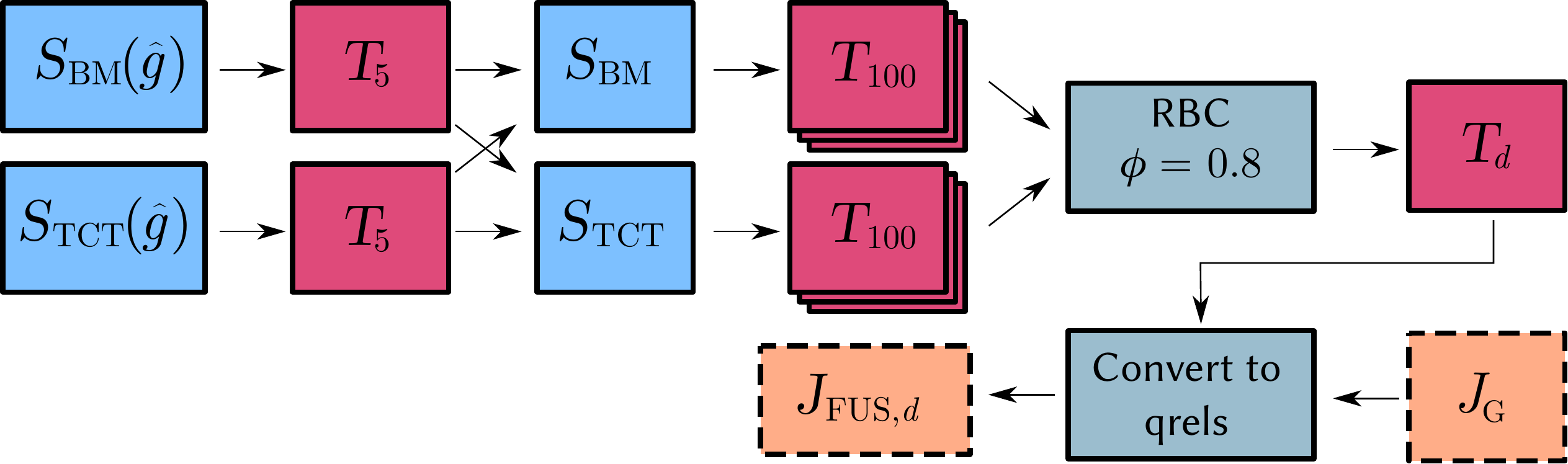}
  \mycaption{Applying rank fusion.
  The top five passages from each of the BM25 and TCT
  query-by-passage runs ($10$ passages) are used as queries to both
  BM25 and TCT.
  The resultant $20$ runs (to depth $100$) are then fused using RBC
  {\citep{bmst17-sigir}}, and the top-$d$ passages of that final run
  are deemed relevant and joined with $\Gol{}(q)$.
  \label{fig-fuse-pipeline}
  }
\end{figure}

The experimental pipeline takes the $\Sys{BM}$ and $\Sys{TCT}$ runs,
together with $\Gol{}(q)$ and one gold passage $\gol{}\in\Gol{}(q)$,
and generates three sets of variable-size {\emph{extrapolated qrels}}:

\begin{itemize}

  \item $\Jud{BM,\,$d$}$ contains $\Gol{}(q)$ plus exactly $d$ additional
  ``deemed relevant'' passages generated via the BM25-based
  query-by-passage process, see Figure~\ref{fig-basic-pipeline}:
  \[
  	\Jud{BM,\,$d$}(q) =
		\Gol{}(q) ~ \cup ~
			\Top{$d$}(\Sys{BM}(\gol{}(q))
				\,\setminus\, \Gol{}(q)) \, .
   \]

  \item $\Jud{TCT,\,$d$}$ is derived from $\Sys{TCT}$ in the same way.

  \item $\Jud{FUS,\,$d$}$ makes use of ten BM25 runs and ten TCT
  runs, with those query-by-passage runs in turn based on two
  original query-by-passage runs, one from each system.
  The fusion process applies the rank-biased centroid approach
  {\citep{bmst17-sigir}} to those twenty runs to obtain a single
  merged run; finally, $d$ top passages are taken from it and added
  to $\Gol{}(q)$, see Figure~\ref{fig-fuse-pipeline}.

\end{itemize}
Those three extrapolated qrels sets, parameterized by the augmentation parameter
$d$, are employed in the experiments described in the next section,
along with the original gold judgments $\Jud{G}$.
Note again that for the majority of queries, $|\Jud{G}(q)|=1$.
In contrast, $|\Jud{BM,\,$d$}(q)| = |\Jud{TCT,\,$d$}(q)| =
|\Jud{FUS,\,$d$}(q)| = |\Jud{G}(q)|+d$.

 \section{Experiments}
\label{sec-experiments}

\myparagraph{Experimental Setup}

We make use of the {\marco} Passage Ranking Collection (version~1).
The {\emph{dev set}} contains qrels for $6{,}980$ queries, with
$6{,}590$ ($94.4$\%) having a single positive label
($|\Gol{}(q)|=1$).
Of the other $390$ ($5.6\%$) queries, $331$ have two labels, $51$
have three labels, and $8$ have four labels.
There are no negative (non-relevant) labels provided in the {\marco} qrels.

\begin{figure}[t]
  \centering
  \includegraphics[width=0.85\textwidth]{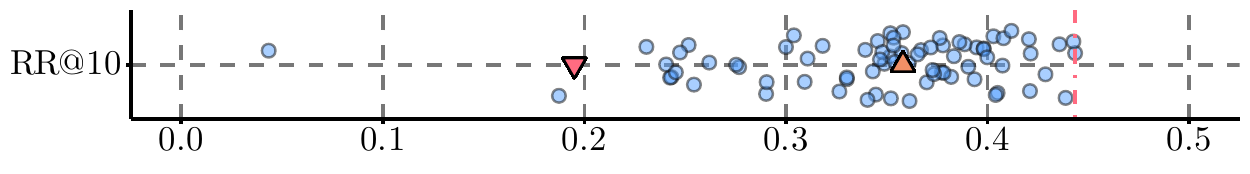}
  \mycaption{Average system scores for RR@10 across the dev set, with
  triangles marking the BM25 (left) and TCT systems (right), and the
  red dot-dashed line representing the best dev run on the official
  leader board as of 24 August 2021.
\label{fig-rr-dist}}
\end{figure}

A total of $75$ system dev runs were used, truncated to $10$ passages
for each query, and with effectiveness computed ``@10'' in all cases.
The runs were a mix of ones that we generated ourselves, and runs
provided by the {\marco} chairs.
Figure~\ref{fig-rr-dist} shows the distribution of system average
(over queries) RR@10 scores.
The two runs used to form the extrapolated qrels were not included in
the~$75$.

\myparagraph{Score Sensitivity}

\begin{figure}[t]
  \centering
  \includegraphics[width=0.9\textwidth]{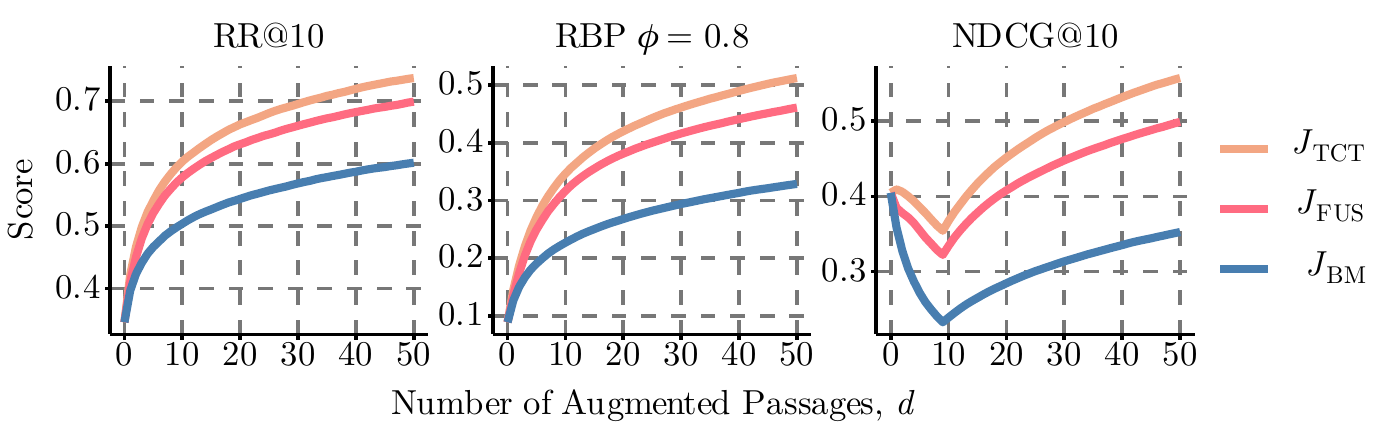}
  \mycaption{Effectiveness scores, averaged across $75$ system runs
  and the dev query set (that is, $75 \times 6980$ values) as
  a function of $d$, the number of further passages deemed
  relevant, for three metrics and three sets of extrapolated
  qrels.
  The original $\Gol{}$-only metric scores correspond to $d=0$.
\label{fig-scores}}
\end{figure}

Figure~\ref{fig-scores} shows how metric scores are affected as
additional ``deemed relevant'' passages are added into the qrels in
a controlled manner.
Unsurprisingly, all three metrics have upward trends, with the
distinctive behavior of NDCG@10 in the vicinity of $d=10$ a
consequence of the normalization process it employs.
The $\Jud{TCT}$ judgments result in the highest average system/query
scores; while the $\Jud{BM}$ judgments give the least score increase.

\myparagraph{Inter-System Sensitivity}

The more important question is whether adding judgments -- in this
case, extrapolated ones -- alters system relativities.
In this experiment, the $\Gol{}$-induced reference ordering of the
$75$ systems is compared with the orderings generated using the
``plus $d$'' extrapolated judgment sets.
Unweighted {\cite{k38-bio}} and top-weighted Kendall's $\tau$
coefficients were computed, in the latter case with a weight of
$1/(k+1)$ assigned to the system at rank $k$ {\cite{v15-www,
s98-spl}}.

\begin{figure}[t]
  \centering
  \includegraphics[width=0.9\textwidth]{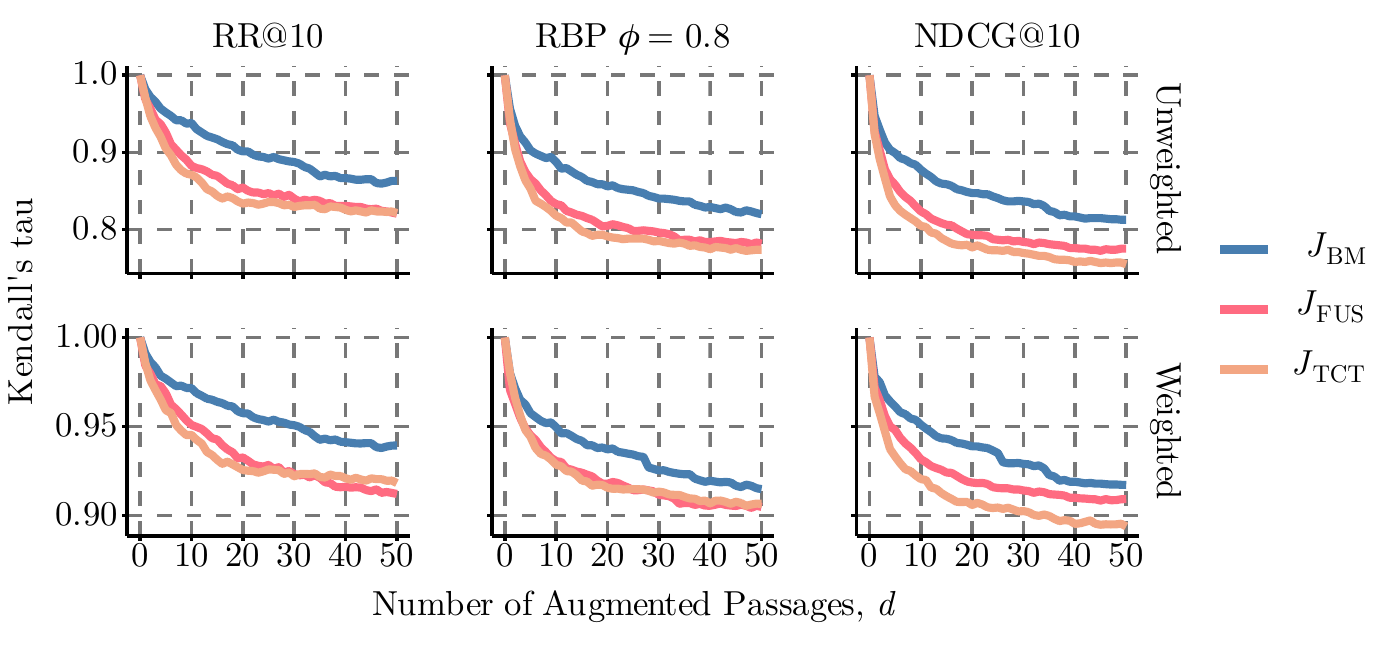}
  \mycaption{Unweighted (top) and weighted (bottom) Kendall's $\tau$
  correlations for 75 systems, all measured relative to the reference
  ordering computed using $\Gol{}$.
  Three different sets of extrapolated judgments are used, and three metrics.
  \label{fig-kendall}}
\end{figure}

\begin{figure}[t]
\centering
\fbox{
\begin{minipage}{0.9\textwidth}\raggedright\scriptsize
  {\bf{Query:}} idaho definition of signed
  \smallskip

  {\bf{Passage 1:}}
  of the state of Idaho and to the administrative jurisdiction of the
  Idaho real estate com - mission, and shall be subject to all
  penalties and remedies available under Idaho law for any violation
  of this chapter.

  \smallskip

  {\bf{Passage 2:}}
  July 2017 Idaho Real Estate License Law \& Rules - i IDAHO REAL
  ESTATE COMMISSION STAFF (208) 334-3285 Administration MiChell M.\
  Bird - Executive Director michell.bird@irec.idaho.gov.....ext.
  105 Jessica Valerio - Administrative Assistant 2

\end{minipage}}
 \caption{A query and two passages,
the first from $\Gol{}$, the second
from $\Jud{FUS}$.
After discussion we judged the second passage to ``answer the query
to approximately the same or greater accuracy'' as the first.
In absolute terms, neither is helpful.
\label{fig-badexample}}
\end{figure}

Figure~\ref{fig-kendall} provides results.
As increasing numbers of plausible passages are deemed to be
relevant, the system orderings tend to slowly diverge from the
reference ordering.
But the top-weighted $\tau$ scores (the lower row) for all three
effectiveness metrics remain above $0.9$, even at $d=20$, indicating
high consistency in the relative performance of the better-scoring
systems.
The $\Jud{BM}$ approach gives the highest $\tau$ values, perhaps
because it disrupts the metric scores the least.

\myparagraph{Judgment Validation}

To provide a limited-scale validation of the extrapolation method,
three passages were extracted from $\Jud{FUS}$ for each of twenty
queries, those at ranks $1$, $2$, and $10$.
Those sixty passages were then judged by each of the authors, and
discussed to reach consensus where we disagreed.
The question considered in all cases was whether the added passage
``answered the query to approximately the same or greater accuracy
than the first passage'', that is, we used the gold passage
$\gol{}(q)$ as an {\emph{anchor}}.
One such pair is shown in Figure~\ref{fig-badexample}; in this
particular example, neither the anchor passage nor the extrapolated
one are relevant, but nor is the second passage less relevant than
the first.

\begin{table}[t]
\centering
\caption{Limited-scale additional judgments as a demonstration of concept.
\label{tbl-annotate}}
\aftercapspace

\renewcommand{\tabcolsep}{1em}
\begin{tabular}{lccc}
\toprule
        & $d=1$ & $d=2$ & $d=10$
\\
\midrule
Fraction judged to be ``as relevant as $\gol{}(q)$''
        & 20/20 & 16/20 & 14/20
\\
\bottomrule
\end{tabular}
 \end{table}

The fraction at each depth $d$ for which the consensus answer was
``yes'' is shown in Table~\ref{tbl-annotate}.
Of the twenty $d=1$ passages, $19$ were simply the gold passage,
$\gol{}(q)$, confirming that in most cases
$\Top{1}(\Sys{}(\gol{}(q))) = \gol{}(q)$.
On the other hand, the results for $d=2$ and $d=10$ provide
compelling preliminary evidence that there are many more relevant
passages in the {\marco} collection than are captured by the
reference qrels, supporting the claims made by {\citet{q+21-naacl}}.

 \section{Conclusion}
\label{sec-conclusion}

We have explored the sensitivity of the {\marco} collection,
measuring the extent to which system scores and system orderings are
stable if more than one passage per query is assumed to be relevant.
Our results show that scores themselves increase as positive qrels
are added, but that system orderings are comparatively resilient.
These findings add credibility to the process used to construct the
{\marco} passage collection.

As a final comment, we observe that effective training of neural
retrieval systems requires negative examples as well as positive ones
{\citep{q+21-naacl,lyl21-rl4nlp}}, a question that has been
considered by a range of authors
{\citep{cjc19-ictir,xx+21-iclr,zm+21-sigir,mdc17-www}}.
Developing an effective mechanism for determining documents or
passages that are plausible answers, but are non-relevant -- as
distinct from ones that are patently non-relevant -- is thus another
interesting challenge.

\myparagraph{Acknowledgements}
This work was supported by the Australian Research Council's
{\emph{Discovery Projects}} Scheme (DP200103136). We thank Nick Craswell and
Bhaskar Mitra for their assistance with the {\marco} dev runs.

\begin{small}
\setlength{\bibsep}{1.2pt}
\bibliographystyle{abbrvnat}

\end{small}

\end{document}